**PROCEEDINGS**

# Search for universal minimum drag resistance underwater vehicle hull using CFD.

## Harsh Vardhan[1*] and Janos Sztipanovits[2]

[1,2] Vanderbilt University, Nashville,37235, USA

*Corresponding Author: Harsh Vardhan. Email: harsh.vardhan@vanderbilt.edu

**ABSTRACT**

In Autonomous Underwater Vehicles (AUVs) design, hull resistance is an important factor in determining the power requirements and range of vehicle and consequently affect battery size, weight, and volume requirement of the design. In this paper, we leverage on AI based optimization algorithm along with Computational Fluid Dynamics (CFD) simulation to study the optimal hull design that minimizing the resistance. By running the CFD based optimization at different operating velocity and turbulence intensity, we want to study/search the possibility of a universal design that will provide least resistance/ near optimal design across all operating condition (operating velocity) and environmental conditions (turbulence intensity). Early result demonstrated that the optimal design found at low velocity and low turbulence condition performs very poor at high velocity and high turbulence conditions. However, design that is optimal at high velocity and high turbulence condition performs near optimal across many considered velocity and turbulence conditions.

**Problem Formulation and Approach:**

**Problem Formulation and Approach:**

AI has proven its potential in system design and operation [1,2,3]. In this work, we want to leverage AI for finding a universal minimum drag hull shape for AUV. First, we formulate the AUV hull design problem as an optimization problem. Let the hull shape indicated as $\Lambda$ is defined using a multivariate parameter $x$, and $f$ is the function that maps a given 3D shape $\Lambda$ with a drag force ($F\_d$) ($f: \Lambda \to F\_d$ ). If $DS$ is the design space of search, the optimization problem can be formulated as

$$\Lambda^* = \underset{x \in DS}{argmin} f(x)$$

The optimization involves running the CFD analysis on chosen hull shape. Since running these CFD simulations are very time-consuming and computationally costly, we chose AI-based Bayesian optimization [4,5,6,7] as our optimization algorithm in a loop with CFD analysis to find the optimal design.

Second, we want to study/search for a universal optimal design that is near optimal across all environmental and operating conditions. For creating the different environment and operating conditions for AUV, we consider five different environmental conditions (turbulence intensity of flow {0.1,2,5,10, 20} percent of mean flow velocity) and five different operating conditions (velocity of AUV {1,2.5,5,7.5,10} in meters/second). The operating velocity and turbulence intensity values are taken from empirical ranges of underwater vehicles' operation. Accordingly, the cartesian product of these two sets creates 25 different scenarios with specific operating velocities and turbulence intensity. By running Bayesian optimization for every 25 scenarios with CFD in the loop, we want to find the optimal hull design in all scenarios.

For running CFD, we chose an open-source simulation tool called openFoam [8] that solves Reynolds-Averaged-Navier Stokes (RANS)[9] with $\kappa - \omega$ Shear Stress Transport ($\kappa - \omega$ SST)[10] as the physical model. The Navies strokes equation is defined below:

$$\rho \frac{d\vec{v}}{dt} = -\nabla p + \mu \nabla^2 v + \rho \vec{g}$$

$$\nabla \cdot \vec{v} = 0$$

For turbulence modeling, we used an improved version of the two-equation turbulence model called $\kappa - \omega$ SST definer by Menter et al.[10] Apart from solving the two equations, the $\kappa - \omega$ SST involves blending functions that ensure gradual changing the $\kappa - \omega$ model of Wilcox to $\kappa - \varepsilon$ model of Jones and Launder from the inner region of a boundary layer to the outer wake region. The two-equation used in $\kappa - \omega$ SST is defined below [11]:

$$\frac{\partial(\rho k)}{\partial t} + \frac{\partial(\rho U_i k)}{\partial x_i} = P_k - \beta^* \rho k \omega + \frac{\partial}{\partial x_i}\left[(\mu + \sigma_k \mu_t)\frac{\partial k}{\partial x_i}\right]$$

$$\frac{\partial(\rho \omega)}{\partial t} + \frac{\partial(\rho U_i \omega)}{\partial x_i} = \alpha \rho S^2 - \beta \rho \omega^2 + \frac{\partial}{\partial x_i}\left[(\mu + \sigma_\omega \mu_t)\frac{\partial \omega}{\partial x_i}\right] + 2(1 - F_1)\rho \sigma_{w2}\frac{1}{\omega}\frac{\partial k}{\partial x_i}\frac{\partial \omega}{\partial x_i}$$

The turbulence intensity ($I$) and mean flow velocity ($U$) define the initial turbulence energy ($k$) and its dissipation rate ($\omega$) during the simulation process. These are related by the following equations:

$$k = \frac{3}{2}(UI)^2$$

$$I = \frac{u'}{U}$$

$$u' = \sqrt{\frac{1}{3}\left({u'_x}^2 + {u'_y}^2 + {u'_z}^2\right)}$$

$$U = \sqrt{U_x^2 + U_y^2 + U_z^2}$$

$$\omega = \frac{k^{\frac{1}{2}}}{l * C_\mu^{\frac{1}{4}}}$$

Here $C_\mu$ is the turbulence model constant which usually takes the value 0.09, $\kappa$ is the turbulent energy, and $l$ is the turbulent length scale. For experimentation, we consider an axisymmetric body of revolution with a fixed hull length (1 meter) and fixed fineness ratio (5) that poses a constraint on the hull diameter, which is 0.2 meters. The rest of the shape of the hull can be changed by 6 control points across the body by changing its location in space. For each of the 25 experiments, the initial turbulence energy and dissipation rate are provided as initial conditions. The hull has two parts -nose and tail and their lengths can be controlled by the parameter nose_length and the tail_length (tail_length=1-nose_length (in meters)). By changing the length of the nose, we can move the fineness ratio constraint across the body and provide an optimizer capability to search for a wide variety of shapes. The deep learning-based method has proven potential to capture the complex physical phenomenon, but these are very data-hungry models [12,13]. Accordingly, for the optimization algorithm, we chose Bayesian optimization, since it is a very sample efficient optimization framework involving expensive function. For modeling our expensive function, we use Gaussian Process non-parametric machine learning architecture to create a probabilistic model of our function ($f$) and Lower Confidence Bound (LCB) as our acquisition function. The Gaussian process [14] is defined as:

$$p(f) = \mathcal{GP}(f; \mu, K)$$

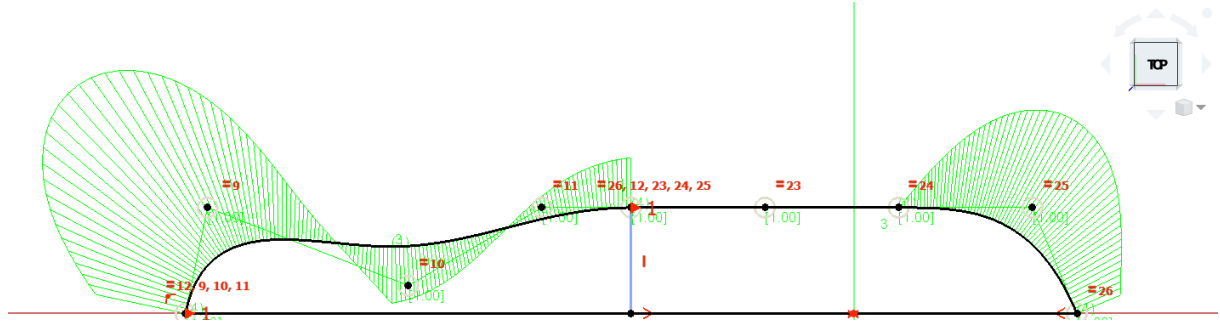

**Fig.1.CAD sketch of parametric hull in FreeCAD[15]**

For any given pair of input points $x, x' \in R^d$, the mean ($\mu$) and covariance ($K$) are defined as

$$\mu(x) = \mathbb{E}(f(x))$$

$$K = \mathbb{E}[(f(x) - \mu(x))(f(x') - \mu(x'))]$$

We selected Lower Confidence Bound (LCB) as our acquisition function ($a$) that measures the utility of candidate point in a Bayesian sequential optimization setting since it has good empirical performance and theoretical guarantee to converge to global optimum and is defined below:

$$a_{LCB}(x; \beta) = \mu(x) - \beta\sigma(x) \; ; \; \beta \geq 0 \; (hyper\ parameter)$$

$$\sigma(x) = \sqrt{K(x,x)}$$

If $f'$ is global optimum and $f(x)$ is the drag of the evaluated sample, a regret function is defined which measure how much have we lost due to selecting a non-optimal sample:

$$r(x) = f' - f(x)$$

$$min \sum_t^T r(x_t) = max \sum_t^T f(x_t) \; ; \; T : budget\ of\ evaluation$$

Srinivas et al [16] proved that with a specific value of $\beta$, the regret will go to 0 asymptotically.

$$\beta = \sqrt{(v, \tau_t)}; v = 1 \text{ and } T_t = 2log(t^{d/2+2}\pi^2/3\delta)$$

$$lim_{T \to \infty} R_T/T = 0$$

We used BO-LCB as our optimizer since it is computationally expensive to evaluate each hull using CFD, and BO is a sample efficient sequential design strategy for global optimization[17] and can find the optimal design in a lesser number of evaluations. The design space (DS) of the search is in 7 dimensions ($\mathbb{R}^7$): six control points with minimum and maximum values in the range are 0 and 0.2 meters respectively for every control point. The seventh dimension of design space is the nose_length whose range is between 10cm-900cm. The tail_length is a derived parameter and is defined as 1000cm-nose_length. We ran 100 iterations of sequential BO for each scenario to find the optimal design. The optimal hull shape and their drag resistances are shown in figure 2 and figure 3 respectively.

**Preliminary Results:**

At every 25 scenarios, we ran our optimization algorithm until convergence or up to the allocated budget of 100 iterations. At end of the optimization process, the discovered optimal designs are shown in Fig 2 and the related drag force is shown in Fig 3. The preliminary results show that at high turbulence intensity, the optimization converged to a single design (rightmost column of fig 2). For comparing the performance of optimal designs, in this work, we consider two optimal designs first, the design obtained at lowest turbulence and lowest speed (D1: vel:1m/s and turbulence 0.1%) and second, the design obtained at highest turbulence and highest speed (D2: vel:10m/s and turbulence 20%)). Once we tested both these designs' performance in all 25 scenarios, we observed, that design D1 produced more drag in all 24 scenarios while design D2 produced significantly lesser drag resistance in 18 out of 24 scenarios (refer to Fig.4 for drag resistance values).

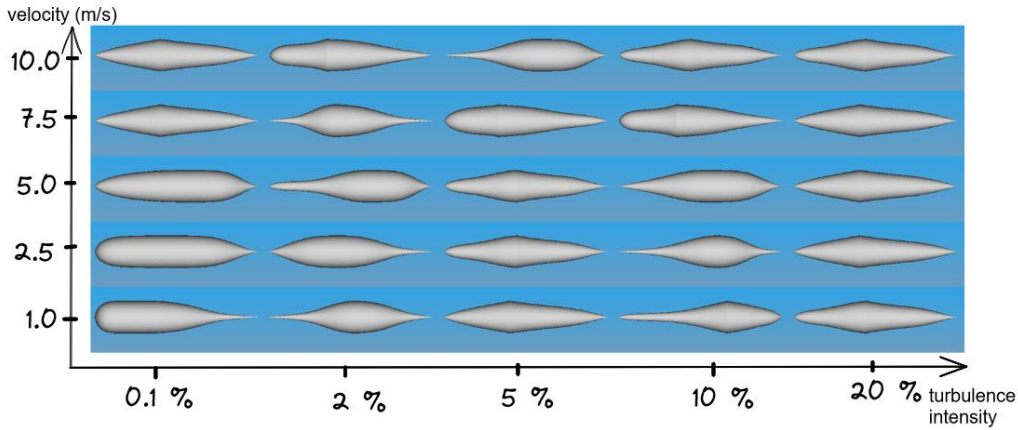

**Fig.2. The optimal hull shapes at each scenarios after optimization.**

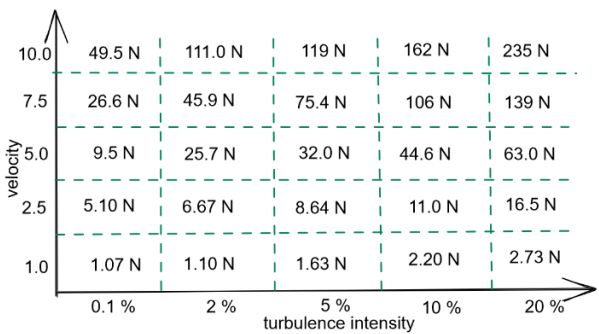

Fig.3.The baseline drag resistances.

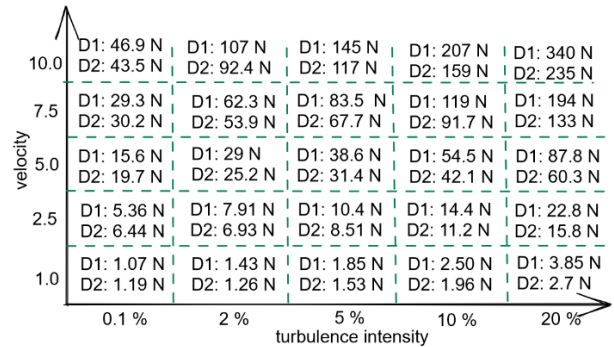

Fig.4.The drag resistances of D1 & D2

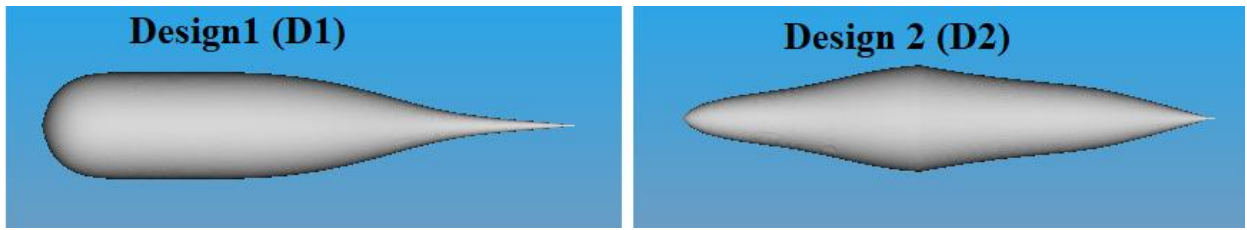

Also, the drag resistance of design D1 increases significantly at high velocity and high turbulence conditions in comparison to design D2. AI is revolutionizing the engineering design of controllers [18] to classical optimization problems [19-22] by designing a controller just by data or solving complex nonlinear optimization in a sample-efficient manner. In this work, we attempted to find a universal minimum drag hull shape by leveraging computational simulation and AI base optimization methods. The future direction is to conclude this research by analyzing each 25 optimal design and their performance across all considered environmental and operating conditions.

**Funding Statement:** This work is supported by DARPA through contract number FA8750-20-C-0537.

**Conflicts of Interest:** The authors declare that they have no conflicts of interest to report regarding the present study.